# Improving the resolution of microscope by deconvolution after dense scan

Yaohua Xie, Yaohua.Xie@hotmail.com

**Abstract** Super-resolution microscopes (such as STED) illuminate samples with a tiny spot, and achieve very high resolution. But structures smaller than the spot cannot be resolved in this way. Therefore, we propose a technique to solve this problem. It is termed "Deconvolution after Dense Scan (DDS)". First, a preprocessing stage is introduced to eliminate the optical uncertainty of the peripheral areas around the sample's ROI (Region of Interest). Then, the ROI is scanned densely together with its peripheral areas. Finally, the high resolution image is recovered by deconvolution. The proposed technique does not need to modify the apparatus much, and is mainly performed by algorithm. Simulation experiments show that the technique can further improve the resolution of super-resolution microscopes.

By now, existing super-resolution techniques have not only surpassed the diffraction-limit, but also improved resolutions significantly[1]. For example, stimulated emission depletion (STED)[2] microscope can illuminate a sample at a spot much smaller than the diffraction limit, e.g., about 10 nm. In this case, the emission property at the targeted location (illuminated spot) can be got by measuring the light intensity. Thereby, the whole image can be got by illuminating and measuring the sample spot by spot. But structures smaller than the spot cannot be resolved in this way. In order to further improve the resolution, we propose a technique termed "Deconvolution after Dense Scan (DDS)". First, it requires some preprocessing on the sample before observation. Then, the sample is scanned densely with conventional approach. Finally, deconvolution[3] is performed on the resulting (blurred) image to get a sharp image with higher resolution. Deconvolution is often used to remove the out-of-focus background in fluorescence microscopy[4], but it is used to further improve the resolution in this study.

In this study, the region need to be observed is called Region of Interest (ROI). The ROIs' shapes can be arbitrary, but a recommended way is to include all the ROIs in a bigger rectangular ROI.

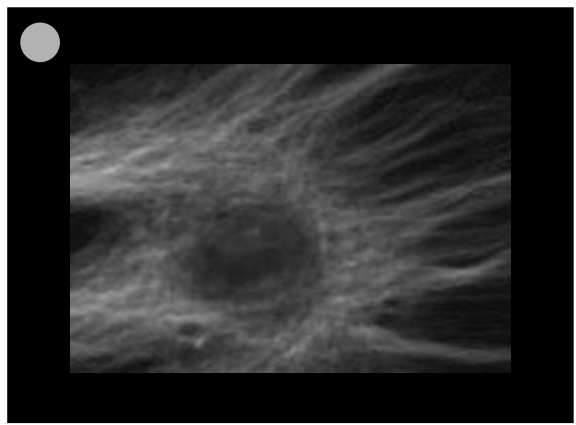

Fig.1. An example of the pre-processing setup

The first step of the proposed method is the pre-processing of sample. It is used to eliminate the optical uncertainty of the peripheral areas around the ROI. For example, the optical characteristics of the sample or background should be known in the peripheral areas around the ROI. In different applications, the optical characteristics may mean light reflection or fluorescence excitation characteristics. Usually, the peripheral areas are comprised of the locations outside the ROI, and the distance between each location and the ROI should be no greater than the size of the illumination spot. One easy way to eliminate the uncertainty is: the light reflection and fluorescence excitation is zero (or ignorable) for everything in the peripheral areas. Fig. 1 shows an example of the pre-processing setup. The ROI is rectangular, and it is surrounded by material with zero reflection and excitation. The illumination spot never beyond the outer border of the material as long as it overlaps (partly or fully) with the ROI. Since the sample is scanned step by step, the pre-processing can be performed for different parts of the ROI at different time. Without this pre-processing, the following deconvolution will not get correct results.

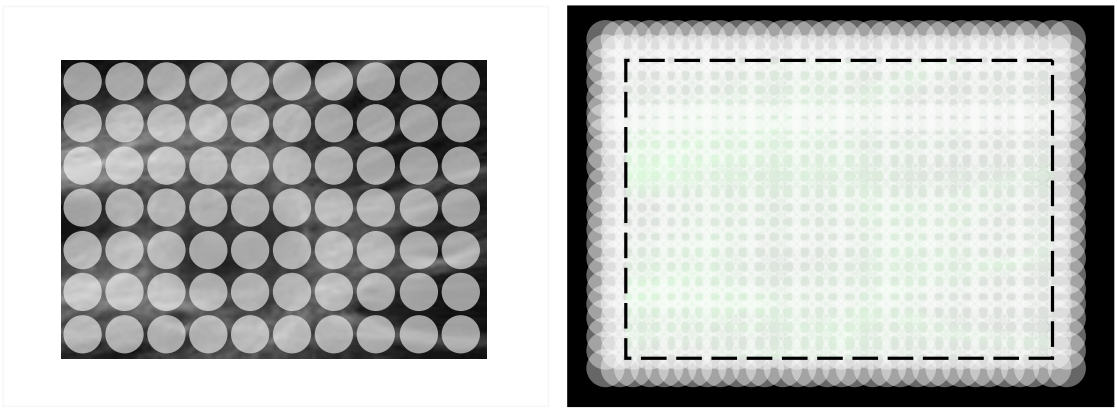

Fig.2. (a) Footprint in usual case　　　　　　　　　(b) Footprint in dense-scan case

Existing techniques, such as STED, illuminates the sample with spot smaller than the diffraction-limit. The spot illuminates different tiny areas in different time. Then, a pixel value is figured out which represents the optical characteristic of each tiny area. Finally, all the pixel values are combined to form a whole image. These techniques can get very high resolution because the spot's size can be much smaller than the diffraction-limit. Fig. 2(a) shows an example of how the spot's "footprints" cover the sample. In the example, the scanning step equals the spot's size. If the step is shorter, adjacent footprints may overlap with one another, as shown by Fig. 2(b). Thereby, the pixels also represent the optical characteristics of overlapped areas. As a result, the combined image looks blurred. But it has more pixels than the usual case because the pixel number is relevant to the scanning steps. Fig. 3(a) and (b) show the combined images of the usual case and the dense-scan case, respectively. Structures smaller the spot's size can be distinguished in none of the two cases.

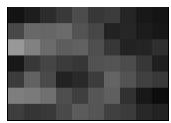　　　　　　　　　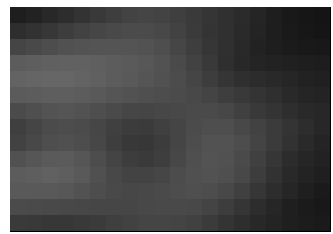

Fig.3. (a) Combined image in usual case　　　　　　　　　(b) Combined image in dense-scan case

In this study, the densely scanned region includes not only the ROI, but also the peripheral areas. If the spot's size is as small as the step length in the dense-scan case, the combined image will be sharp. Such an image is the high resolution image we want, and is termed "expected image" here. But the spot is actually larger than the step. So let's treat the spot's light intensity distribution as an image, and term it "spot image". It is essentially a matrix whose elements represent the light intensity values at different locations (grids). When the expected image is convolved with the spot image, the result would be a blurred image. It can be proved that the blurred image is exactly the combined image Fig. 3(b), in the condition created by the pre-processing. In short, the blurred image Fig. 3(b) is the convolution of the expected image with the spot image.

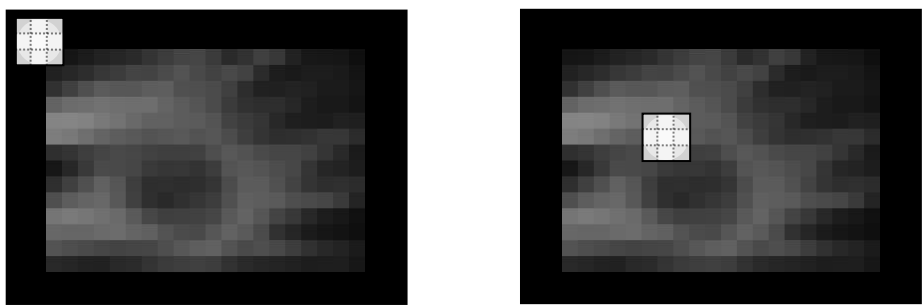

Fig.4. (a) Spot at the peripheral areas　　　　　(b) Spot at the ROI area

The convolution is illustrated in Fig. 4. The center of the spot image lies in the peripheral areas in Fig. 4(a), and lies in the ROI in Fig. 4(b). No matter in which cases, an Airy-disk-shaped pattern of the illuminated area (spot) can be get by the microscope. Then the total intensity of the reflected light or excited fluorescent can be figured out. This value is used as a pixel in the combined (blurred) image. According to principles of optics [5], reflected light or excited fluorescent is approximately equal (or proportional) to "the product of illumination and sample's optical characteristic". Multiply the spot image with the illuminated image pixel-wise, and then sum all the values of the resulting image (matrix). The sum will be equal (or proportional) to the pixel in the blurred image. Thereby, the convolution result is actually the blurred image; or, it can be get by multiplying the latter by a coefficient. When the peripheral areas' light reflection and fluorescence excitation is zero, the corresponding pixel values in the expected image are also zeros. This guarantees that the above conclusion also correct when the spot lies at the image's boundary, as shown by Fig. 4(a). If the peripheral material's optical characteristic is not zero, it should be known and introduced into the form of convolution. In that case, the convolution would be more complicated, but the procedure is totally similar. If the reflected light or excited fluorescent is more complicated than "the product of illumination and sample's optical characteristic", similar procedure can also be employed to build the relationship between images, and then the expected image can also be recovered with corresponding approaches.

The above convolution does not happen by nature or in a computer program, but performed through a manual procedure instead. Thereby, it can be called "manual convolution" or "conceptual convolution". Since the blurred image is only used to recover the expected image, it is termed "intermediate image" here. The spot image is acquired in advance and used as another input to the deconvolution. Any feasible approaches can be employed to perform the deconvolution, e.g., inverse filtering, Weiner filtering, solving a system of equations, blind deconvolution, optimization, the combination of multiple approaches, etc.

In the convolution, the spot image (function of light intensity distribution) plays a role of Point Spread Function (PSF). But it has a significant difference from the PSF of conventional light microscope. The spot PSF has zero values at all the locations except a central area. After an image is convolved with it, the result still has complete components including high frequency. On the contrary, a microscope's PSF extends infinitely, and is equivalent to a low pass filter in Fourier domain[6]. It removes all the high frequency components by convolution. In order to improve efficiency, multiple spots could be used for simultaneous scan as long as they do not affect one another. After deconvolution, the recovered image is sharp and has more pixels than that in usual case. In other words, it has higher resolution and includes more detailed structures.

Simulation experiments are performed to test the proposed technique. Fig. 5 shows the result of a typical experiment. In this experiment, the sample is simulated by an image, i.e., the expected image. A physical sample actually has unlimited details, but the expected image only need to include sharp details at the expected resolution. In this experiment, the expected resolution, i.e., the resolution of the expected image is 0.1 nm/pixel. That means the sample should be scanned with a step of 0.1 nm. The ROI is as large as the expected image which has 300x300 pixels, i.e., 30nm x 30nm. Fig. 5(a) shows the expected image, and Fig. 5(b) shows the result when it is observed by a conventional microscope. The simulated microscope has an Airy disk whose radius is about 200 nm. That is about 2000 pixels in the expected resolution. Fig. 5(b) looks very blurred because it is the convolution of the expected image with the microscope's PSF. The simulated spot 's diameter is 10.1 nm, thereby the spot image is 101x101 pixels in the expected resolution. In usual scan case, the spot's footprints are side by side. Only one pixel can be figured out for each footprint (illuminated location). In other words, the resolution is 10.1 nm/pixel, and thereby the combined image has less than 3x3 pixels (300/101 = 2.9). Almost no details can be distinguished in such a small image. On the contrary, the combined image has 500x500 pixels in the dense-scan case, as shown by Fig. 5(c). Please note that it covers both the ROI and the peripheral areas, and is termed "intermediate image". Fig. 5(d) shows the result recovered by the proposed method. It is sharper than the intermediate image, and much sharper than the conventional result. The averaged difference of pixel is 1.3565e-11 between the recovered image and the expected image.

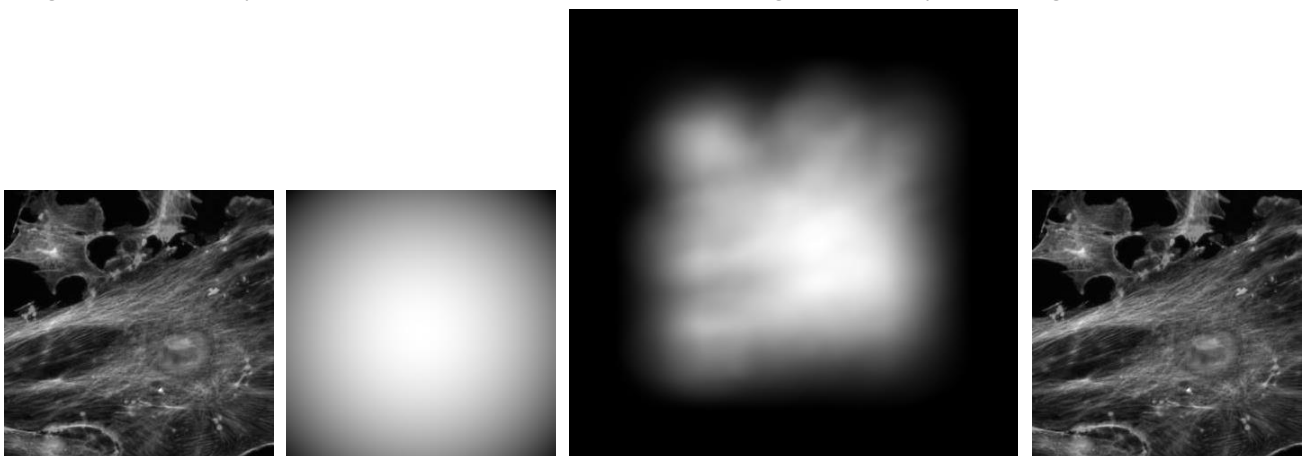

Fig.5. (a) Expected image     (b) Conventional result     (c) Intermediate image     (d) Recovered image

As demonstrated by the experiments, the proposed technique can further improve the resolution of existing super-resolution microscopes. Furthermore, it does not require to change apparatus much, and most the job can be done by the algorithm.

But there are still some barriers in practice. For example, it could be difficult to get the spot image accurate enough. Noises in any one of the images/data could distort the recovered results. Therefore, it is important to investigate and solve these problems in the future.

**References**
1. Sigal, Y.M., R.B. Zhou, and X.W. Zhuang, *Visualizing and discovering cellular structures with super-resolution microscopy.* Science, 2018. **361**(6405): p. 880-887.
2. Hell, S.W. and J. Wichmann, *Breaking the diffraction resolution limit by stimulated emission: stimulated-emission-depletion fluorescence microscopy.* Optics Letters, 1994. **19**(11): p. 780-782.
3. Gonzalez, R.C. and R.E. Woods, *Digital Image Processing (3rd Edition)*. 2014: PEARSON.
4. ZENG, Z.-p., et al., *Computational methods in super-resolution microscopy.* Frontiers of Information Technology & Electronic Engineering, 2017. **18**(09): p. 1222-1236.
5. Born, M., et al., *Principles of Optics: Electromagnetic Theory of Propagation, Interference and Diffraction of Light*. 1999: Cambridge University Press.
6. Goodman, J.W., *Introduction to Fourier Optics*. 2017: W. H. Freeman.